\documentclass[12pt]{article}
\usepackage{epsfig,psfrag,cite,float}
\def \beq{\begin{equation}}
\def \eeq{\end{equation}}
\def \bea{\begin{eqnarray}}
\def \eea{\end{eqnarray}}
\def \ben{\begin{enumerate}}
\def \een{\end{enumerate}}
\def \bit{\begin{itemize}}
\def \eit{\end{itemize}}

\def \branch{{\cal B}}
\def \eff{\hbox{eff}}
\def \Im{{\hbox{Im}}\,}
\def \Re{{\hbox{Re}}\,}
\def \gev{{\hbox{GeV}}}

\def \tev{{\hbox{TeV}}}

\def \cl#1{{#1\%\ \mathrm{C.L.}}}

\def \fig#1{Fig.~\ref{#1}}

\def \nn{\nonumber}
\def \rf{Ref.~\cite}

\def \b{\beta}

\def \g{\gamma}

\def \d{\delta}

\def \m{\mu}

\def \s{\sigma}

\begin{document}
\pagestyle{empty}

\begin{center} 
{\bf \large
Improved Model-Independent Analysis of Semileptonic \\ 
and Radiative Rare $B$ Decays}

\bigskip

Enrico Lunghi\footnote{E-mail address: lunghi@mail.desy.de}\footnote{
This work is supported by the Swiss National Fond}

{\it 
Insitut f\"ur Theoretische Physik, Univerit\"at Z\"urich,
8057 Z\"urich, Switzerland}
\end{center}

\begin{abstract}
We study the impact of recent $B$-factories measurements and upper
limit of radiative and semileptonic rare $B$-decays. We present model
independent constraints on the relevant Wilson coefficients and show
the impact on the parameter space of some concrete realizations of the
Minimal Supersymmetric Standard Model.
\end{abstract}

\section{Experimental inputs}
In this talk we use the most recent experimental results on inclusive
and exclusive $b\to s \g$ and $b\to s \ell^+\ell^-$ decays to update
the analysis presented in Ref.~\cite{aghl}. The experimental results
that we use in the analysis are
\bea
\branch (B\to X_s \g) &=& (3.40^{+0.42}_{-0.37}) \times 10^{-4} 
\hbox{~\cite{aleph,cleo,babarbsg,bellebsg}}, \label{bsgexp}\\
\branch (B\to K \ell^+ \ell^-) &=& (0.63^{+0.14}_{-0.13})\times 10^{-6} 
\hbox{~\cite{babar,belle}}, \label{bkllexp} \\
\branch (B\to K^* \mu^+ \mu^-) &\leq& 3.0 \times  10^{-6} \; {\rm at}\; \cl{90} 
\hbox{~\cite{babar,belle}}, \label{bksmmexp}  \\
\branch (B\to K^* e^+ e^-) &=&(1.68^{+0.68}_{-0.58}\pm 0.28)\times 10^{-6} 
\hbox{~\cite{babar}}, \label{bkseeexp} \\ 
\branch (B\to X_s \mu^+ \mu^-) &= & (7.9\pm2.1^{+2.0}_{-1.5})\times  10^{-6} 
\hbox{~\cite{belleincl}}, \label{bsmmexp} \\
\branch (B\to X_s e^+ e^-) &=& (5.0\pm 2.3^{+1.2}_{-1.1}) \times 10^{-6} 
\hbox{~\cite{belleincl}}, \label{bseeexp} \\
\branch (B\to X_s \ell^+ \ell^-) &=& (6.1\pm 1.4^{+1.3}_{-1.1}) \times 10^{-6} 
\hbox{~\cite{belleincl}}. \label{bsllexp} 
\eea
The $B\to X_s \g$ branching ratio is given with a cut on the photon
energy ($E_\g > m_b/20$) and the value presented is the weighted
average of the four available measurements. The branching ratios for
the semileptonic modes (\ref{bkllexp}--\ref{bsllexp}) refer to the
non--resonant branching ratios integrated over the dilepton invariant
mass spectrum.  On the theoretical side, this amounts to consider only
the perturbative part of the amplitude and to leave out all the
resonant contributions (that are usually shaped using Breit-Wigner
ans\"atze). On the experimental one, judicious cuts are used to remove
the dominant resonant contributions arising from intermediate $c\bar
c$ resonances ($J/\psi,\psi^\prime,...$); SM-based theoretical
distributions~\cite{ali-res} are then used to correct the data for the
experimental acceptance. Finally, let us stress that the $B\to X_s e^+
e^-$ branching ratio is given with a cut on the di-lepton invariant
mass, $M_{ee}\equiv \sqrt{s} > 0.2\; \gev$, in order to remove virtual
photon contributions and the $\pi^0\to ee\g$ photon conversion
background.  The $B\to X_s e^+e^-$ rate increases steeply for $s \to
0$ (because of the almost real photon pole) and is extremely sensitive
to the Wilson coefficient of the magnetic moment operator (that
controls the $b\to s \gamma$ transition). The use of the branching
ratio without extrapolation to the full $s$ spectrum, allows to reduce
the uncertainties in the comparison with the Standard Model (SM)
prediction and to properly analyze models with an enhanced magnetic
moment Wilson coefficient.

\section{Theoretical framework and SM predictions}
The effective Hamiltonian in the SM inducing the $b\to s \ell^+
\ell^-$ and $b\to s\g$ transitions is (up to negligible contributions
proportional to $V_{us}^* V_{ub}^{}$)
\begin{eqnarray}
    \label{Heff}
    {\cal H}_{\eff} =  - \frac{4G_F}{\sqrt{2}} V_{ts}^* V_{tb}^{}
    \sum_{i=1}^{10} C_i(\mu) \, O_i (\mu)\quad ,
\end{eqnarray}
where $G_F$ is the Fermi constant and $V_{ij}^{}$ are the CKM matrix
elements. $O_i(\mu)$ are dimension-six operators at the scale $\mu$
and $C_i(\mu)$ are the corresponding Wilson coefficients. The most
relevant operators are (the complete list can be found, for instance,
in~\rf{aghl})
\bea
\label{oper}
    O_7    & = & \frac{e}{g_s^2} m_b (\bar{s}_{L} \sigma^{\mu\nu} b_{R}) F_{\mu\nu} \, , \\
    O_8    & = & \frac{1}{g_s} m_b (\bar{s}_{L} \sigma^{\mu\nu} T^a b_{R}) G_{\mu\nu}^a \, , \\ 
    O_9    & = & \frac{e^2}{g_s^2}(\bar{s}_L\gamma_{\mu} b_L)
                 \sum_\ell(\bar{\ell}\gamma^{\mu}\ell) \, , \\
    O_{10} & = & \frac{e^2}{g_s^2}(\bar{s}_L\gamma_{\mu} b_L)
                 \sum_\ell(\bar{\ell}\gamma^{\mu} \gamma_{5} \ell) \, , 
\eea
where the subscripts $L$ and $R$ refer to left- and right- handed
components of the fermion fields. 

The differential decay width for the inclusive decay $B\to X_s
\ell^+\ell^-$ is given by the parton level result supplement by
calculable power corrections. In the NNLO approximation, the
non--resonant decay width can be written as
\bea
\label{eq:rarewidthpower}
 && \hskip -0.6cm
    \frac{d\Gamma(b\to s \ell^+\ell^-)}{d\hat s} =
    \left(\frac{\alpha_{em}}{4\pi}\right)^2
    \frac{G_F^2 m_{b,pole}^5\left|V_{ts}^*V_{tb}^{}\right|^2}
    {48\pi^3}(1-\hat s)^2 
    \left[ \left (1+2\hat s\right)
    \left (\left |\widetilde C_9^{\eff}\right |^2+
    \left |\widetilde C_{10}^{\eff}\right |^2 \right ) G_1( \hat s) 
\right. \nn \\
 && \hskip +2.8cm
    + \left. 4\left(1+2/\hat s\right)\left
    |\widetilde C_7^{\eff}\right |^2 G_2( \hat s) +
    12 \mbox{Re}\left (\widetilde C_7^{\eff}
    \widetilde C_9^{\eff*}\right ) G_3( \hat s) + G_c(\hat s) \right] \; ,
\eea
where $\hat s \equiv s/m_b^2$ and the functions $G_i (\hat s)\;
(i=1,2,3)$ and $G_c (\hat s)$ encode respectively the $1/m_b^2$ and
$1/m_c^2$ corrections. $\widetilde C_i^{\eff}$ are effective Wilson
coefficients (whose explicit form is given in \rf{aghl}) that are
functions of the dilepton mass squared and incorporate part of the
operator matrix elements. In particular, $\widetilde C_9^{\eff}$
contains contributions due to perturbative $c\bar c$ rescattering and
develops an imaginary part for $s > 4 m_c^2$. Performing the $\hat s$
integration and constraining $s_{ee} > (0.2 \;\gev)^2$, the decay
widths in electrons and muons are essentially equal and are given by
the following numerical formula:
\bea
 && \hskip -0.75cm \branch (B\to X_s \ell^+ \ell^-) = \Big[
    4.534 
    + 8.665 \; |C_7^{\rm tot}|^2 
    + .119 \; (|C_9^{\rm NP}|^2 + |C_{10}^{\rm NP}|^2) 
    + .996 \; \Re  C_7^{\rm tot} C_9^{\rm NP*} \nn \\
 && \hskip -0.8cm + 4.130 \;  \Re C_7^{\rm tot} 
    + 0.171 \; \Im  C_7^{\rm tot} 
    + 1.068 \; \Re C_9^{\rm NP} 
    + .064 \; \Im C_9^{\rm NP} 
    -1.011 \; \Re C_{10}^{\rm NP} \Big] \times 10^{-6} \, , \nn
\eea
where $C_9^{\rm NP}$ and $C_{10}^{\rm NP}$ are the new physics
contributions to $C_9(\m_W)$ and $C_{10}(\m_W)$ evaluated at $\m_W
\simeq m_W$ and $C_7^{\rm tot}$ is the sum of the SM ($C_7^{\rm SM}
(\m_b)$) and new physics ($C_7^{\rm NP} (\m_b)$) contributions
evaluated at $\m_b\simeq 2.5 \; \gev$. A detailed discussion of all
the assumptions that enter this formula can be found in \rf{aghl}.  In
the SM we find
\beq
\label{bsll}
\branch(B\to X_s \ell^+ \ell^-) = 
\left(4.15 \pm 0.27\pm 0.21\pm 0.62 \right) \times 10^{-6}
= \left(4.15 \pm 0.70 \right) \times 10^{-6}
\eeq
where the errors correspond to variations of $\mu_b$, $m_{t,pole}$ and
$m_c/m_b$. Comparing this estimate with the experimental measurements
(\ref{bsmmexp})--(\ref{bsllexp}) we see that there is agreement with
the SM at the 1 $\s$ level. Note that Previous experimental data on
the di-electron channel were extrapolated using SM-based distributions
and that the branching ratio for $B\to X_s e^+ e^-$ integrated over
the whole dilepton invariant mass spectrum is~\cite{aghl}
$(6.89\pm1.01)\times 10^{-6}$ (i.e. the $s_{ee} < (0.2 \; \gev)^2$
region enhances (\ref{bsll}) by $66 \%$). This clearly shows to what
extent the choice to give branching ratios not extrapolated allows for
a cleaner identification of new physics effects.

For what concerns the exclusive decays $B\to K^{(*)} \ell^+ \ell^-$,
we implement the NNLO corrections calculated by Bobeth et al.~in
\rf{BMU} and by Asatrian et al.~in \rf{AAGW} for the short-distance
contribution.  Then, we use the form factors calculated with the help
of the QCD sum rules in \rf{Ali:2000mm}. For lack of space we can not
describe some subtleties related to the treatment of the so--called
hard spectator interactions and to the value of the magnetic moment
form factor at $s=0$ (see \rf{aghl} for a complete discussion). 
The SM NNLO predictions that we obtain are
\bea
\branch (B\to K \ell^+ \ell^-) &=& (0.35 \pm 0.12) \times 10^{-6} \;\;\; 
        (\d \branch_{K\ell\ell} = \pm 34 \%) \; , \\
\branch (B\to K^* e^+ e^-) &=& (1.58 \pm 0.49) \times 10^{-6} \;\;\; 
        (\d \branch_{K^*ee} = \pm 31 \%) \; , \\
\branch (B\to K^* \m^+ \m^-) &=& (1.19 \pm 0.39) \times 10^{-6} \;\;\; 
        (\d \branch_{K^*\m\m} = \pm 33 \%) \; .
\eea
From the comparison with Eqs.~(\ref{bkllexp})--(\ref{bksmmexp}) we see
that in the $B\to K \ell^+\ell^-$ channel there is a 1.6 $\s$
discrepancy with the SM expectation while in the $K^*$ channels there
is a perfect agreement.

\section{Model independent analysis}
\begin{figure}[t]
\begin{center}
\epsfig{file=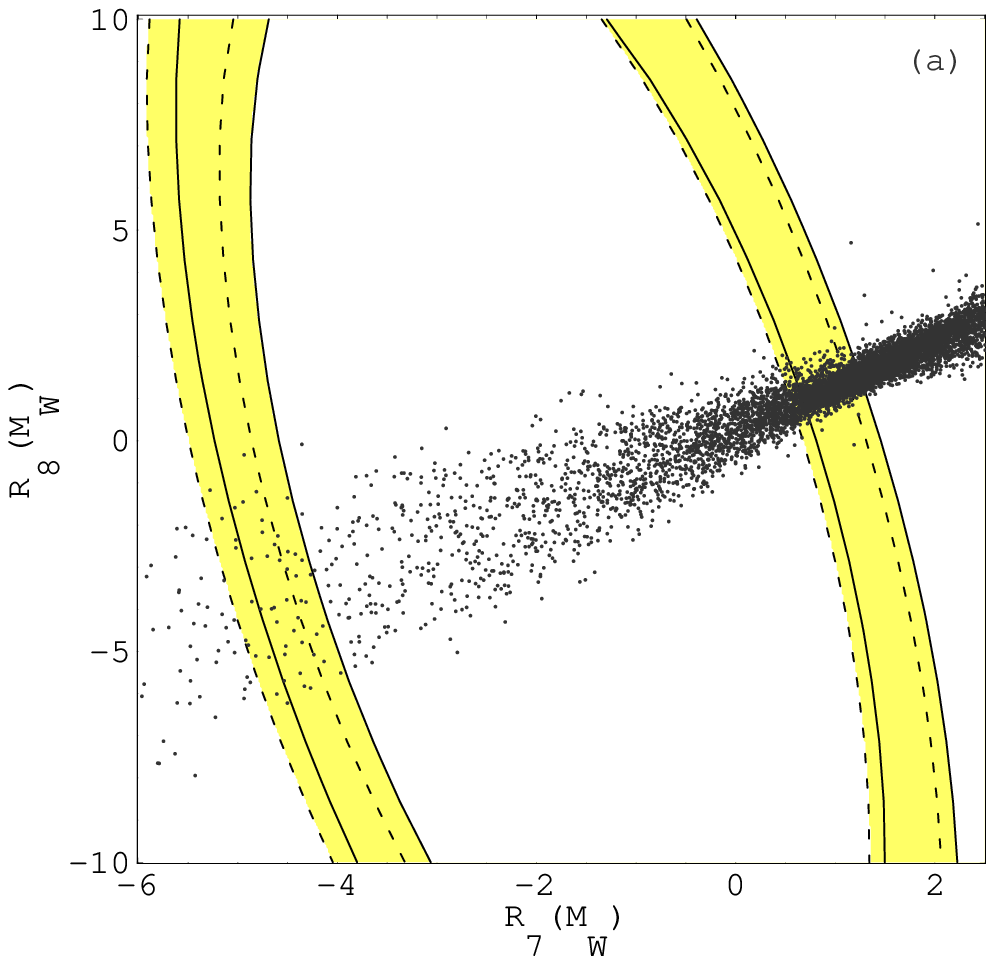,width=0.48\linewidth}
\hspace*{.2cm}
\epsfig{file=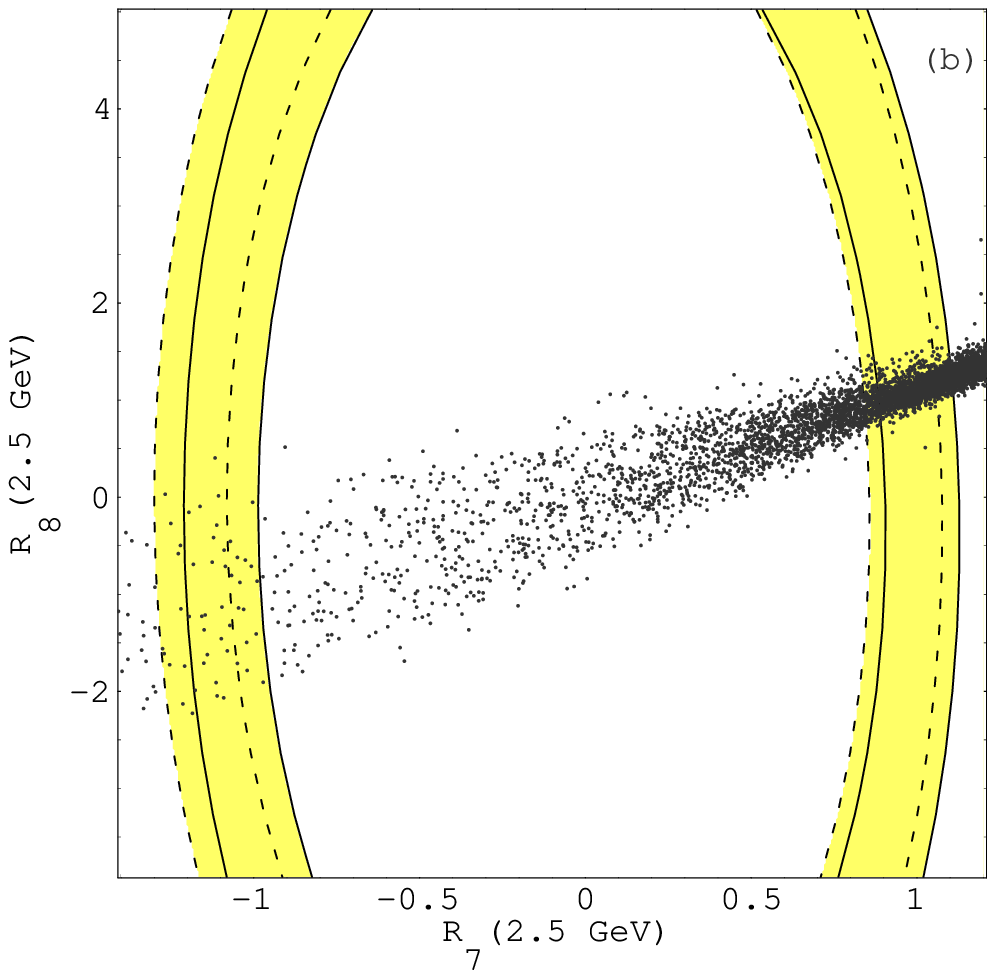,width=0.48\linewidth}
\caption{\it $\cl{90}$ bounds in the $[R_7 (\m), R_8(\m)]$ plane
following from the world average $B\to X_s \g$ branching ratio for
$\mu=m_W$ (left-hand plot) and $\mu=2.5$ GeV (right-hand plot).
Theoretical uncertainties are taken into account. The solid and dashed
lines correspond to the $m_c = m_{c,pole}$ and $m_c =
m_{c}^{\overline{MS}} (\m_b)$ cases respectively. The scatter points
correspond to the expectation in MFV models.}
\label{bsg}
\end{center}
\end{figure}

The first step consists in extracting the bounds that the measurement
(\ref{bsgexp}) implies for $C_7^{\rm tot} (2.5 \; \gev)$. The main
difficulty arises from the treatment of the $m_c$ dependence of the
$B\to X_s \g$ branching ratio. In \rf{Gambino:2001ew}, it was noted
that, in this decay, the charm quark mass enters the matrix elements
at the two-loop level only and that it would be more appropriate to
use the running charm mass evaluated at the $\m_b \simeq O(m_b)$
scale, leading to $m_c/m_b \simeq 0.22\pm 0.04$, compared to
$m_{c,pole}/m_b \simeq 0.29\pm 0.02$. This is a reasonable choice
since the charm quark enters only as virtual particle running inside
loops; formally, on the other hand, it is also clear that the
difference between the results obtained by interpreting $m_c$ as the
pole mass or the running mass is formally a NNLO effect. In what
concerns $b \to s \ell^+ \ell^-$, the situation is somewhat different,
as the charm quark mass enters in this case also in some one-loop
matrix elements. In these one-loop contributions, $m_c$ has the
meaning of the pole mass when using the expressions derived in
Ref.~\cite{AAGW}. Since the bounds on the $C_7$ do not depend
dramatically on $m_c$, we just derive them using both values of the
charm mass and taking the union of the allowed ranges. We present the
results of this analysis in Figs.~\ref{bsg}a and \ref{bsg}b, where we
show the allowed regions in the $R_7$ and $R_8$ plane obtained using
the $\cl{90}$ $B\to X_s \g$ bound (here $R_{7,8}\equiv C_{7,8}^{\rm
  tot} / C_{7,8}^{\rm SM}$). We take $|R_8(\mu_W)| \leq 10$ in order
to satisfy the constraints from the decays $b\to s g$ and $B\to X_{c
  \!\!  /}$~\cite{GL}.The regions in \fig{bsg}b translate in the
following allowed constraints:
\bea
\cases{
m_c/m_b = 0.29:  \;\;\; C_7^{\rm tot} (2.5 \; \gev)\in [-0.37,-0.18] 
                                  \; \& \; [0.21,0.40] \; , & \cr 
m_c/m_b = 0.22:  \;\;\; C_7^{\rm tot} (2.5 \; \gev)\in [-0.35,-0.17] 
                                  \; \& \; [0.25,0.43] \; . & \cr}
\eea
In the subsequent numerical analysis we impose the union of the above
allowed ranges
\bea
\label{a7lim}
 -0.37 \leq C_7^{\rm tot,<0} (2.5 \; \gev) \leq -0.17 
  & \& &
  0.21 \leq C_7^{\rm tot,>0} (2.5 \; \gev) \leq 0.43 
\eea
calling them $C_7^{\rm tot}$--positive and $C_7^{\rm tot}$--negative
solutions.

\begin{figure}[t]
\begin{center}
\epsfig{file=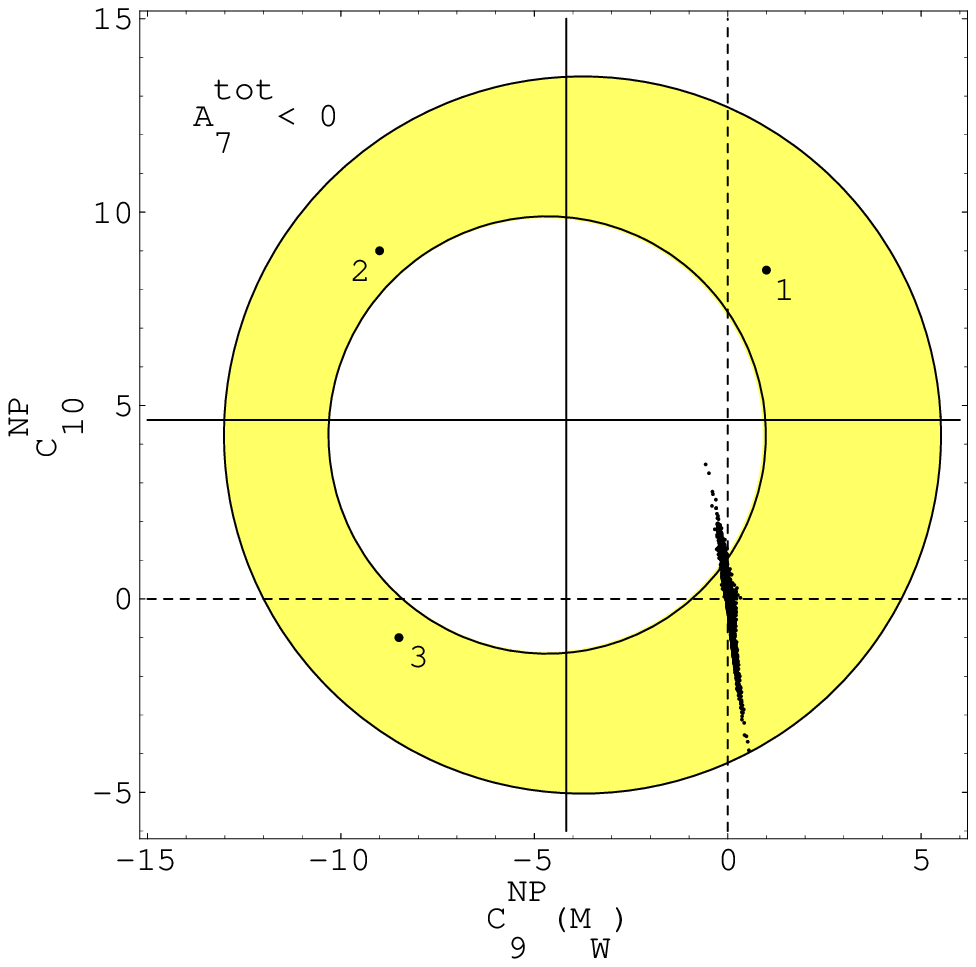,width=0.4\linewidth}
\hskip 0.5cm 
\epsfig{file=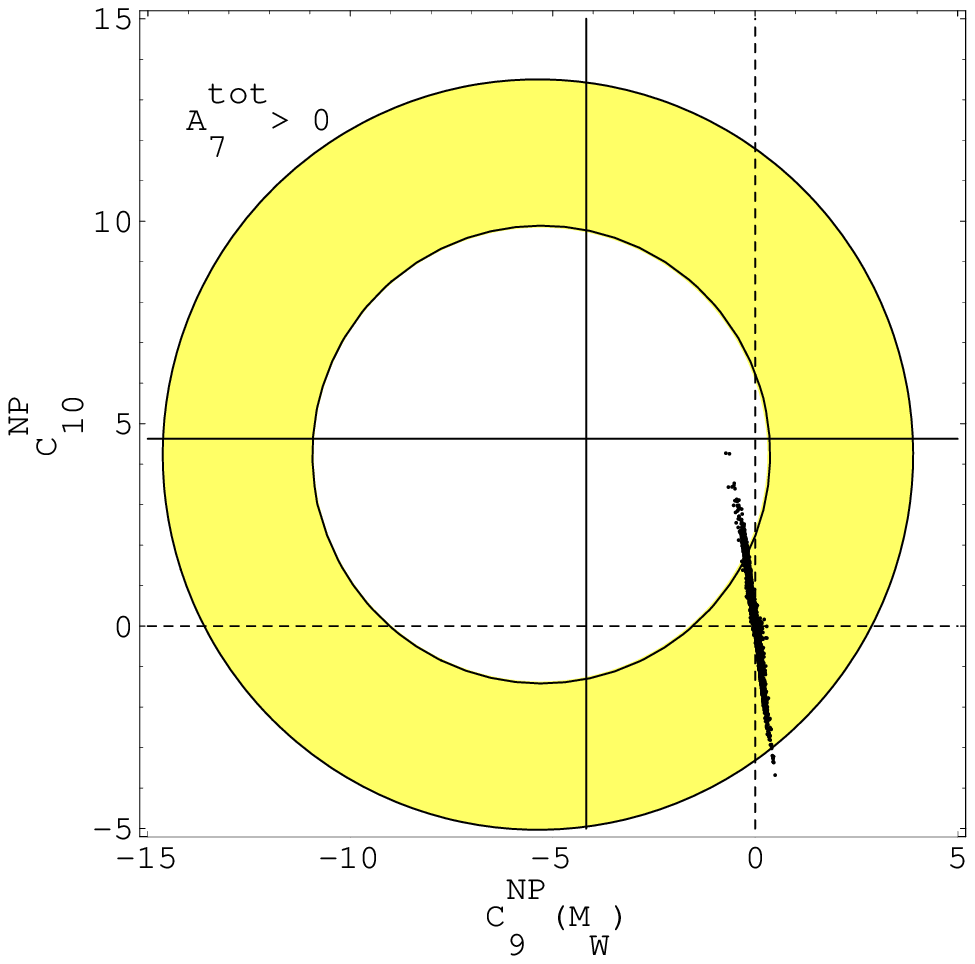,width=0.4\linewidth}
\caption{\it {\bf NNLO Case.} Constraints on new physics contributions 
  to the Wilson coefficients $C_9$ and $C_{10}$ implied by $b\to s
  \ell^+\ell^-$ decays. The plots correspond to the $C_7^{\rm tot}(2.5
  \; \gev)<0$ and $C_7^{\rm tot}(2.5 \; \gev) >0$ case, respectively.
  The points are obtained by means of a scanning over the EMFV
  parameter space and requiring the experimental bound from $B\to X_s
  \g$ to be satisfied.}
\label{fig:total}
\end{center}
\end{figure}

We present the results of the model independent analysis of $b\to s
\ell^+\ell^-$ decays in Fig.~\ref{fig:total}. Within each plot we vary
$C_7^{\rm tot}$ inside the allowed ranges (\ref{a7lim}) and plot the
$\cl{90}$ constraints implied by Eqs.~(\ref{bkllexp})--(\ref{bsllexp})
in the $[C_9^{\rm NP}(\m_W),C_{10}^{\rm NP}]$ plane. The SM correspond
to the point (0,0). In each plot the inner and outer contours are
determined by the measurements of the decays $B\to K \ell^+\ell^-$ and
$B\to X_s \ell^+\ell^-$ respectively.
\section{Analysis in supersymmetry}
In this section we analyze the impact that the measurements
(\ref{bsgexp})--(\ref{bsllexp}) have on three variants of the minimal
supersymmetric standard model (MSSM), namely minimal flavour violation
(MFV), gluino mediated contributions and extended minimal flavour
violation (EMFV). 

\begin{figure}[t]
\begin{center}
\epsfig{file=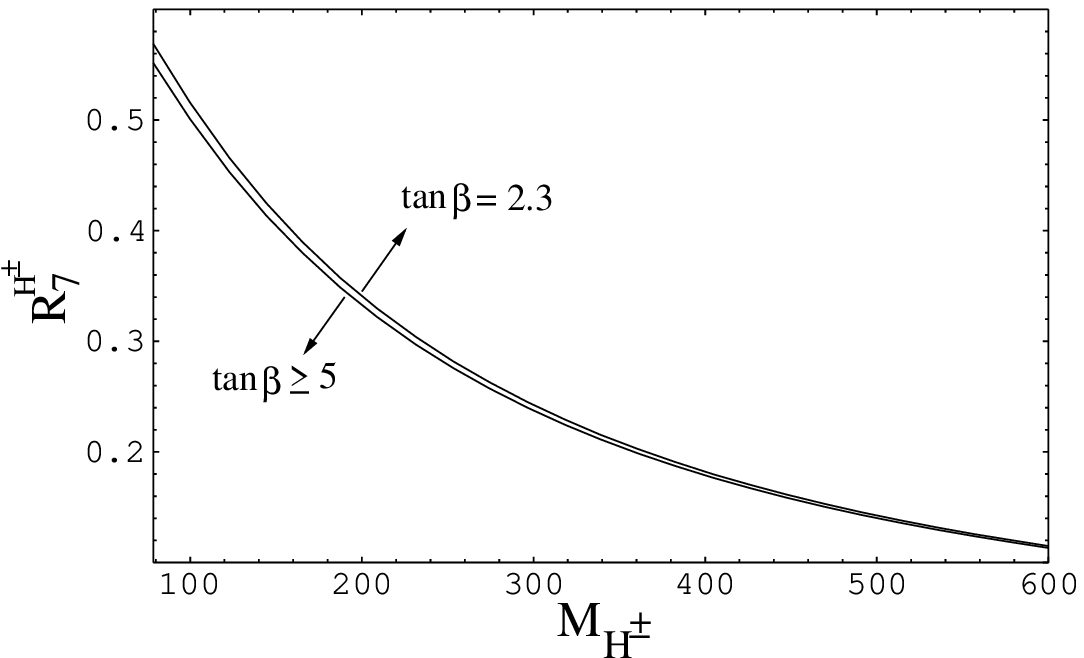,width=0.49\linewidth}
\hfill
\epsfig{file=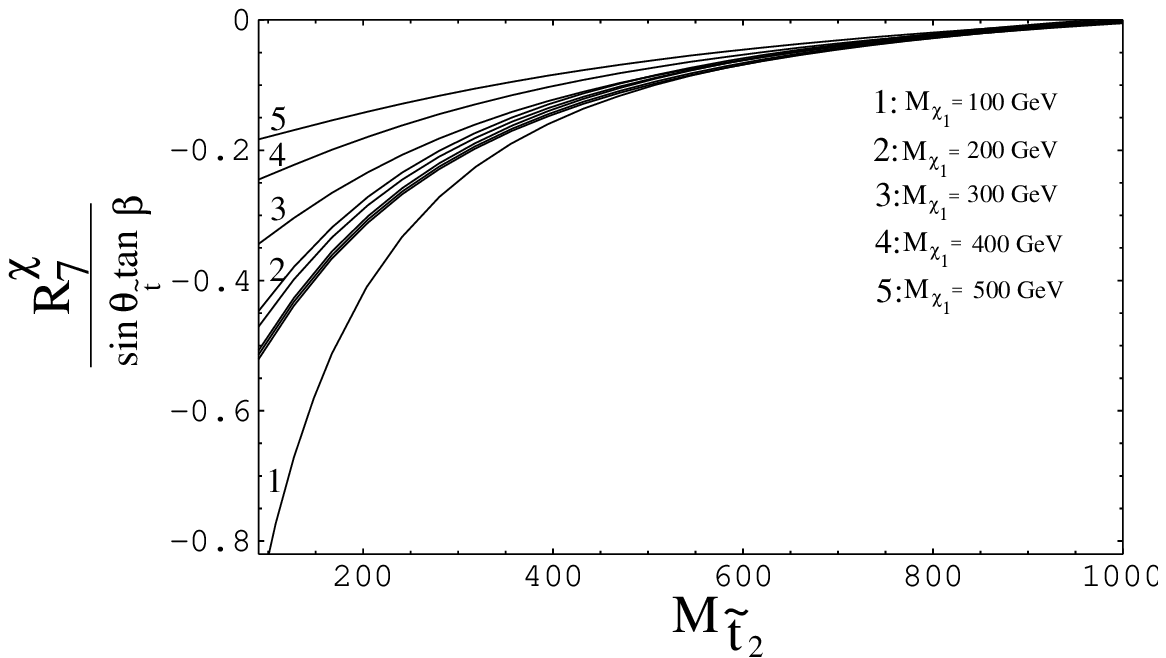,width=0.49\linewidth}
\caption{\it {\bf Left plot:} Dependence of 
  $R_7^{H^\pm} (\m_b) \equiv C_7^{H^\pm} (\m_b) / C_7^{\rm SM} (\m_b)$
  on the mass of the charged Higgs. {\bf Right plot:} Dependence of
  $R_7^{\chi} (\m_b) \equiv C_7^{\chi} (\m_b) / C_7^{\rm SM} (\m_b)$
  on the mass of the lightest stop in MFV models.  The chargino
  contribution is essentially proportional to $\sin \theta_{\tilde t}
  \tan \b$ for not too small $\sin \theta_{\tilde t}$.  For the curve
  2 we show the variation due to several choices of $\theta_{\tilde
    t}$ and $\tan \beta$.}
\label{fig:r7ch}
\end{center}
\end{figure}

{\bf MFV.}  As already known from the existing literature (see
for instance \rf{LMSS}), minimal flavour violating contributions are
generally too small to produce sizable effects on the Wilson
coefficients $C_9$ and $C_{10}$. Indeed, scanning over the MFV parameter 
space and imposing the lower bounds on the sparticle masses we obtain
\beq
C_7^{\rm tot}<0 : \cases{C_9^{MFV}(\m_W) \in [-0.2, 0.4] \cr  
                                   C_{10}^{MFV}    \in [-1.0, 0.7]}\; , \;\;\;\;
C_7^{\rm tot}>0 : \cases{C_9^{MFV}(\m_W) \in [-0.2, 0.3]\, , \cr  
                                   C_{10}^{MFV}    \in [-0.8, 0.5]} \; .
\eeq
From the comparison of the size of these contributions with the
allowed regions depicted in Fig.~\ref{fig:total} we see that the
current experimental results on $b\to s\ell^+\ell^-$ decays are not
precise enough to constraint the MFV parameter space.  The situation
is completely different for what concerns $b\to s\g$. The scatter plot
presented in Fig.~\ref{bsg} is obtained varying the MFV SUSY
parameters and shows the strong correlation between the values of the
Wilson coefficients $C_7$ and $C_8$. In fact, the SUSY contributions
to the magnetic and chromo--magnetic coefficients differ only because
of colour factors and loop-functions. In Fig.~\ref{fig:r7ch} we
present, finally, the dependence of the charged Higgs and chargino
contributions to $C_7$ on the relevant mass parameters (that are the
charged Higgs mass for the former and the lightest chargino and stop
masses for the latter).  From these figures is clear that the
knowledge of the sign of $C_7^{\rm tot}$ will strongly constrain the
MFV parameter space, by putting upper or lower limits on the chargino
and stop masses.

{\bf Gluino contributions.}  Gluino contributions to $C_9$ and
$C_{10}$ are governed by mass insertions in the down squark mass
matrix. From the analysis presented in \rf{LMSS} we see that the
dominant diagrams involve the parameter $(\d^d_{23})_{LL}$ and that
large deviations from the SM are unlikely. 

{\bf Extended--MFV models.} A basically different scenario arises if
chargino--mediated penguin and box diagrams are considered. As can be
inferred by Table~4 in \rf{LMSS}, the presence of a light $\tilde t_2$
generally gives rise to large contributions to $C_9$ and especially to
$C_{10}$. EMFV models are based on the heavy squarks and gluino
assumption.  In this framework, the charged Higgs and the lightest
chargino and stop masses are required to be heavier than $100 \; \gev$
in order to satisfy the lower bounds from direct searches.  The rest
of the SUSY spectrum is assumed to be almost degenerate and heavier
than $1 \; \tev$. The lightest stop is almost right--handed and the
stop mixing angle (which parameterizes the amount of the left-handed
stop $\tilde t_L$ present in the lighter mass eigenstate) turns out to
be of order $O(m_W / M_{\tilde q}) \simeq 10\%$. The assumption of a
heavy ($\ge 1$ TeV) gluino totally suppresses any possible
gluino--mediated SUSY contribution to low energy observables. Note
that even in the presence of a light gluino (i.e. $M_{\tilde g} \simeq
O(300 \; \gev)$) these penguin diagrams remain suppressed due to the
heavy down squarks present in the loop. In the MIA approach, a diagram
can contribute sizeably only if the inserted mass insertions involve
the light stop.  All the other diagrams require necessarily a loop
with at least two heavy ($\geq 1 \; \tev$) squarks and are therefore
automatically suppressed. This leaves us with only two unsuppressed
flavour changing sources other than the CKM matrix, namely the mixings
$\tilde u_L - \tilde t_2$ (denoted by $\d_{\tilde u_L \tilde t_2}$)
and $\tilde c_L - \tilde t_2$ (denoted by $\d_{\tilde c_L \tilde
  t_2}$).  We note that $\d_{\tilde u_L \tilde t_2}$ and $\d_{\tilde
  c_L \tilde t_2}$ are mass insertions extracted from the up--squarks
mass matrix after the diagonalization of the stop system and are
therefore linear combinations of $(\d_{13})^U_{LR}$,
$(\d_{13})^U_{LL}$ and of $(\d_{23})^U_{LR}$, $(\d_{23})^U_{LL}$,
respectively. In \fig{fig:total} we present the results of an high
statistic scanning over the EMFV parameter space requiring each point
to survive the constraints coming from the sparticle masses lower
bounds and $b\to s \g$. Note that these SUSY models can account only
for a small part of the region allowed by the model independent
analysis of current data. In the numerical analysis reported here, we
have used the integrated branching ratios alone to put constraints on
the effective coefficients. This procedure allows multiple solutions,
which can be disentangled from each other only with the help of both
the dilepton mass spectra and the forward-backward asymmetries. Only
such measurements would allow us to determine the exact values and
signs of the Wilson coefficients $C_7$, $C_9$ and $C_{10}$.

\end{document}